\documentclass[
reprint, 
superscriptaddress, 
amsmath,
amssymb, 
aps,
floatfix,
longbibliography,
onecolumn
]{revtex4-1}

\usepackage{enumitem}
\usepackage{float}

\usepackage{soul}
\setstcolor{red}

\usepackage{graphicx}
\usepackage{xcolor}
\usepackage[colorlinks=true,citecolor=blue,linkcolor=magenta]{hyperref}
\usepackage{bm}

\begin{document}

\title[Decoherence factor as a convolution]{Decoherence factor as a convolution: \\an interplay between a Gaussian and an exponential coherence loss}

\author{Bin Yan}
\affiliation{Theoretical Division, Los Alamos National Laboratory, Los Alamos, New Mexico 87545, USA}
\affiliation{Center for Nonlinear Studies, Los Alamos National Laboratory, Los Alamos, New Mexico 87545, USA}
\author{Wojciech H. Zurek}
\affiliation{Theoretical Division, Los Alamos National Laboratory, Los Alamos, New Mexico 87545, USA}

\begin{abstract}
We identify and investigate the origin and nature of the transition between Gaussian and exponential forms of decoherence: The decoherence factor (that controls the time dependence of the  off-diagonal terms of the density matrix expressed in the pointer basis representation) is the convolution of the Fourier transforms of the spectral density and of the overlap (between the eigenstates the environment with and without couplings to the system). Spectral density alone tends to lead to the (approximately) Gaussian decay of coherence while the overlap alone results in a (largely) exponential decay. We show that these two contributions combine as a convolution, their relative importance controlled by the strength of the system-environment coupling. The resulting decoherence factor in the strong and weak coupling limits leads to predominantly Gaussian or exponential decay, respectively, as is demonstrated with two paradigmatic examples of decoherence---a spin-bath model and the quantum Brownian motion.
\end{abstract}

\maketitle

\section{Introduction}

Quantum systems lose quantum coherence when coupled to their environments. Decoherence results in decay of the off-diagonal terms between system's pointer states---states that are immune to decoherence---in its density matrix \cite{Zurek1981,Zurek1991,Schlosshauer2019-tg}. It is an essential ingredient responsible for the quantum-to-classical transition \cite{Zurek2003-qi,Zurek2006-vj} and the main obstacle in the development of quantum information technologies \cite{Buluta2009-xn,Preskill2018-xh,Nielsen2010-gq}. 

Decoherence is caused by the  flow of quantum information from the system $\cal S$ into its environment $\cal E$. Decoherence turns the initial superpositions of pointer states into their mixture. Decoherence factor is often presumed to decay exponentially, but this is not always the case. For instance, power law decay may appear~\cite{Fiori2006-ng,Schomerus2007-vt,Cai2013-ci,Beau2017Nonexponential,Pekola2020-jm,Sarkar2017Nonexponential}. Nonetheless, exponential decay is widely regarded as typical in various settings \cite{Lindblad1976Generators,Gorini1976Completely,Benatti1998-ua,Knight1976-ap,Knight1976-jq,Burgarth2017-qq}.

At short times, decoherence takes quadratic forms related to quantum Zeno paradox \cite{vonNeumann, Misra1977-gf, Facchi2008-go,Baunach2021Copycat}. This transient behavior is often quickly replaced by an exponential decay, although sometimes quadratic decay initiates a Gaussian time dependence  \cite{Cucchietti2005-yh,Zurek2007-ny}. 

The interplay between Gaussian and exponential decoherence has not been---as yet---understood or even investigated. 
We show that decoherence factor can be described by a convolution $f*g$
of an exponential $f(t)=e^{-\Gamma |t|/2}$ and a Gaussian $g(t)=e^{-\sigma^2 t^2/2}$:
\begin{eqnarray}\label{eq:intro}
        (f*g)(t)\equiv \int_{-\infty}^{\infty} d\tau f(\tau)g(t-\tau)
\end{eqnarray}%
The Gaussian contribution originates from the spectrum: It is the Fourier transform of the spectral density (density of states) of the interaction Hamiltonian.
Its origin is easiest to understand for pure decoherence, when the self-Hamiltonians of e.g. the spin-$\frac 1 2$ $\cal S$ and the $N$ spin-$\frac 1 2$ $\cal E$ can be neglected~\cite{Cucchietti2005-yh,Zurek2007-ny}. The $2^N$ eigenvalues of the interaction Hamiltonian are then obtained by summing the couplings of $\cal S$ to the environment spins with the sign (+ or -) given by their relative orientations. Even for modest $N$ the distribution of the eigenvalues approaches quickly a Gaussian. Gaussian spectrum arises in more general circumstances for physical systems with local interactions \cite{Kota2014-cg}.

The exponential factor accounts for the response of the environment to the system-environment interactions: According the Fermi's golden rule, the environment decays from its initial state due to the coupling with the system with the rate $\Gamma$ determined by the coupling strength. 

The Fourier transform of the decoherence factor is then simply a product of two Fourier images. One is the energy spectrum;
the other `Breit-Wigner - like' represents the decay of the overlap between the eigenstates of the environment with and without coupling to the system. The overlap function will be defined rigorously in the next section.

These two contributions are responsible for the interplay between exponential and Gaussian time-dependence of the decoherence factor. In the limit of weak ($\Gamma\ll\sigma$) and strong ($\Gamma\sim\sigma$) coupling, the convolution---hence, the decoherence factor---exhibits predominantly exponential or Gaussian decay, respectively. Below we derive the convolution form of the decoherence factor in the spin-bath model, and then apply it to another paradigmatic model of decoherence, the quantum Brownian motion.

\section{Spin-bath model}

The model consists of a single spin linearly interacting with a generic environment. 
The composite $\cal SE$ can be described by the Hamiltonian:
\begin{eqnarray}\label{eq:Hamiltonian}
    H=\lambda \sigma_z\otimes H_\mathcal{I}+H_\mathcal{E},
\end{eqnarray}
where $\sigma_z$ is the Pauli operator, $H_{\mathcal{I},\mathcal{E}}$ act on the environment, and $\lambda$ is a dimensionless coupling strength. 
This Hamiltonian is relevant for a large class of pure decoherence processes of single-spin systems, and has been investigated in various contexts \cite{Zurek1982-un,Braun2001-el,Cucchietti2005-yh,Zurek2007-ny} including quantum phase transitions \cite{Quan2006-pc,Damski2011-bz}. 

Consider the composite $\cal SE$ that starts as a product:
\begin{eqnarray}\label{eq:initialstate}
    \rho_{\mathcal{SE}}(0)=\rho_\mathcal{S}(0)\otimes\rho_\mathcal{E}(0)
\end{eqnarray}
and the central spin is prepared in a pure state $|\psi_\mathcal{S}(0)\rangle=a|0\rangle+b|1\rangle$. For the sake of simplicity, in the following analysis, the environment initial state $\rho_\mathcal{E}$ is assumed to be an eigenstate $|n\rangle$ of $H_\mathcal{E}$ with energy $E_n$. The results can be generalized (e.g., to a thermal state) by averaging with respect to the specific ensemble.

\begin{figure}[t]
    \centering
    \includegraphics[width=\textwidth]{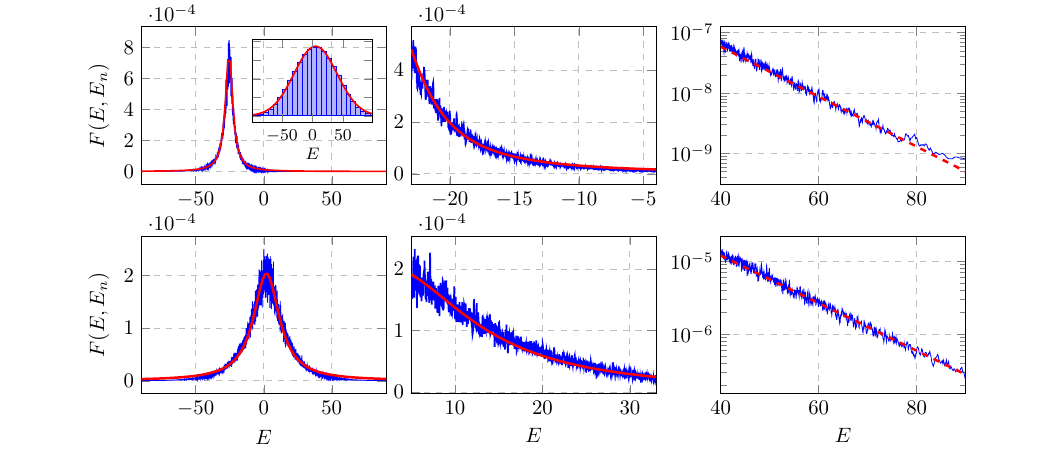}
    \caption{The overlap function of the spin-bath model with 14 bath spins. Upper and lower panels correspond to different initial energy eigenstates and coupling strengths, i.e., $\lambda = 1$ and $2$, respectively. The first column is the plots of the full overlap functions, with a zoom-in shown on the second column and the third column (semi-log scale). Red solid (dashed) curves are the best fit to the Lorentzian (exponential) function. Inset is the histogram for the (Gaussian) spectral density.}
    \label{fig:strength}
\end{figure}

Decoherence of the central spin is reflected in the decoherence factor $r(t)$ that suppresses off-diagonal element of its reduced density matrix $\rho(t)$:
\begin{eqnarray}
  r(t)\equiv \rho_{12}(t)/(ab^*).
\end{eqnarray}
Simple algebra shows that it is given by:
\begin{eqnarray}
            r(t)= \langle n|e^{i( H_\mathcal{E}+\lambda H_\mathcal{I})t}e^{-i( H_\mathcal{E}-\lambda H_\mathcal{I})t}|n\rangle
            \equiv \langle n|e^{i H_\mathcal{E}t}e^{-i( H_\mathcal{E}t +\lambda H_\mathcal{P})}|n\rangle,\label{eq:decoherencefactor0}
\end{eqnarray}
where we have introduced an effective ``perturbation'' $H_\mathcal{P}$ using the Baker-Campbell-Hausdorff formula, which appears in the same order as the coupling $H_\mathcal{I}$ (see Appendix). The second line of the above equation can be interpreted as the survival probability of the initial state after an imperfect echo loop, i.e., a forward evolution followed by a perturbed backward evolution. This quantity is also known as the Loschmidt echo \cite{Cucchietti2003-hh,Goussev2012-ep,Gorin2006-ro}.

Denote the eigenenerges of the ``perturbed'' Hamiltonian, $H_\mathcal{E}+ \lambda H_\mathcal{P}$, as $E^k$, and the corresponding eigenstate as $|k\rangle=\sum_n C^k_n |n\rangle$. The decoherence factor (\ref{eq:decoherencefactor0}) further reduces to a Fourier transformation,
\begin{eqnarray}\label{eq:decoherencefactor}
    r(t) &= e^{iE_nt}\sum_k e^{-iE^kt} |C_n^k|^2 = e^{iE_nt} \int d E \ e^{-iEt} F(E,E_n) \eta(E).
\end{eqnarray}
Here, the discrete sum is replaced by a continuous integral with a ``weighting function''---the global spectral density $\eta(E)$. $F(E^k,E_n) \equiv \overline{|C^k_n|^2}$ is an averaged smooth function in continuous energy $E^k$ of the squared overlap $|C_n^k|^2$ between the perturbed and unperturbed eigenfunctions, which we call the \emph{overlap}. The local density of states (LDOS), a.k.a the strength function, is defined as
\begin{eqnarray}\label{eq:ldos}
    P(E,E_n) \equiv F(E,E_n) \eta(E).
\end{eqnarray}
Both the LDOS and the overlap have peaks around the resonance energy $E_n$. The LDOS was introduced and intensively studied in the field of nuclear physics \cite{Bohr1998-aa} and encountered elsewhere (see e.g. \cite{Gruss2016Landauer}). In Ref.~\cite{Bohr1998-aa}, using random matrix theory, it was derived
to generally take the form
\begin{eqnarray}\label{eq:breitwigner}
    P(E,E_n) =\frac{1}{2\pi} \frac{\Gamma(E)}{(E-E_n)^2+(\Gamma(E)/2)^2},
\end{eqnarray}
where $\Gamma(E)=2\pi V^2\eta(E)$ is given by Fermi's golden rule; $V^2$ is the average of the square of the off-diagonal elements of the perturbation matrix $\lambda H_\mathcal{P}$, when written in the basis of the unperturbed Hamiltonian $H_\mathcal{E}$. It is known \cite{Flambaum2000-ck,Santos2012-qx,Santos2012-lo} that in the weak perturbation regime, where the width of the peak of the LDOS is much smaller than the width of the spectral density $\eta(E)$, $\Gamma(E) \approx 2\pi V^2 \eta(E_n)$ can be treated as a constant. In this case, the LDOS reduces to a Lorentzian (Breit-Winger) distribution, whose Fourier transformation yields an exponential decay in time. 
On the other hand, for large perturbations, the LDOS can be approximated by a Gaussian \cite{Santos2012-qx,Santos2012-lo}, with a width bounded by the bandwidth of the spectral density. 

To provide a unified treatment of the decoherence factor, including the cross-over regime, we first note that the LDOS (\ref{eq:breitwigner})
is valid for an arbitrary perturbation strength \cite{Bohr1998-aa}. Comparing with definition (\ref{eq:ldos}), we identify the overlap function that takes the form
\begin{eqnarray}\label{eq:spread}
    F(E,E_n) = \frac{V^2}{(E-E_n)^2+\left[\pi V^2 \eta(E)\right]^2}.
\end{eqnarray}
This is not a precise Lorentzian distribution since the `width' term is energy-dependent. Nevertheless, we will show that the overlap is insensitive to this energy dependence, and has an approximate (unnormalized) Lorentzian form, even for large perturbations that make the LDOS (\ref{eq:ldos}) far from a Lorentzian shape.

To establish this we perform a Taylor expansion of the energy-dependent width $\Gamma(E)$ around the peak position $E_n$. The LDOS (\ref{eq:breitwigner}) is of a Lorentzian form only at zeroth order of the expansion. However, the overlap $F(E,E_n)$ has an exact (unnormalized) Lorentzian form up to the second order. Both first and second-order energy-dependent terms can be absorbed into the quadratic term $(E-E_n)^2$. In this case, the effective width $\Gamma_{\rm eff}$ of the overlap is
\begin{eqnarray}\label{eq:rate}
    \left(\frac{\Gamma_{\rm eff}}{2}\right)^2 = \frac{2\pi^2 V^4 \eta^2}{2+\pi^2 V^4 {\eta^2}''} - \left[\frac{\pi^2 V^4{\eta^2}'}{2+\pi^2 V^4 {\eta^2}''}\right]^2,
\end{eqnarray}
where $\eta(E)$ and its derivatives are all evaluated at $E_n$. This expansion also shifts the effective peak position of the overlap by a small energy residue, $E_{\rm eff} = E_n - E_r$, where $E_r = (\pi^2 V^4 {\eta^2}')/(2+\pi^2 V^4 {\eta^2}'')$.

The corresponding Fourier transform of the overlap (\ref{eq:spread}) is then proportional to an exponential decay $e^{-\Gamma_{\rm eff} |t|/2}$ multiplied by a phase, which, after absorbing the phase factor $e^{iE_nt}$ in the decoherence factor (\ref{eq:decoherencefactor}), eventually contributes to a net oscillation $e^{iE_rt}$. 

In physical systems, the Hamiltonian is in general given by few-body interactions or local terms, whose spectral densities satisfy universal Gaussian distributions \cite{Kota2014-cg}. This, upon Fourier transform, gives rise to a Gaussian decay in the time domain, $e^{-\sigma^2t^2/2}$, where $\sigma$ is the bandwidth (standard deviation) of the spectral density. Here, without the loss of generality, we assume the spectral density has a maximum at $E_0 = 0$. A non-zero $E_0$ only contributes a phase factor $e^{iE_0t}$ to the decoherence factor.

The decoherence factor (\ref{eq:decoherencefactor}), up to an overall slow oscillation $e^{iE_rt}$, is the Fourier transform of the product of two spectral images in the energy domain---the Lorentzian overlap and the Gaussian spectral density. The Fourier transform of such a product is a convolution of the two individual Fourier transforms of these two spectral images. They are exponential and Gaussian functions in the time domain, as derived above. Consequently, the decoherence factor is
\begin{eqnarray}\label{eq:convolution}
    r(t) \propto \int_{-\infty}^{\infty} d\tau  e^{-\Gamma |\tau|/2}e^{-\sigma^2 (t- \tau)^2/2}
         \propto \sum_{\pm} e^{\pm\Gamma t/2} \rm{Erfc}\left(\frac{\Gamma/2\pm\sigma^2 t}{\sqrt{2}\sigma}\right),
\end{eqnarray}
where Erfc is the complementary error function. The full derivation is presented in the appendix.

For weak ($\Gamma \ll \sigma$) and strong ($\Gamma \sim \sigma$) couplings, this expression reduces to exponential and Gaussian decays, respectively. When we define the decoherence time as the time scale for the $\mathcal{O}(1)$ decay of the decoherence factor, in the exponential and Gaussian regime, the decoherence time scales as $\sim 1/\Gamma$ and $\sim 1/\sigma$, respectively. 

Recall also that $\lambda$ quantifies the interaction strength between the system and the environment. According to Fermi's golden rule, for $\lambda\ll 1$, $\Gamma \propto \lambda^2$. For the other extreme case, when the interaction eventually dominates the environment internal dynamics as studied in Refs.~\cite{Cucchietti2005-yh,Zurek2007-ny}, the bandwidth of the Hamiltonian, $\sigma \propto \lambda$. We conclude that the decoherence time scales as $\sim \lambda^{-2}$ and $\sim \lambda^{-1}$ in the weak and strong coupling limit, respectively. 

It is worth stressing that, the above derivation of the convolution description of the decoherence factor is based on two explicit assumptions, that is, i) The energy spectrum of a typical physical Hamiltonian (with local interactions) takes a universal Gaussian form, and ii) the LDOS take a Lorentzian-like form (\ref{eq:breitwigner}). These two conditions can be derived using random matrix theory \cite{Bohr1998-aa,Kota2014-cg} under mild assumptions, and has been tested in numerous contexts. In this sense, we claim the convolution description of the decoherence factor of the spin-bath model is typical. Hence, for a physical system with local interactions, that can be well described by the spin-bath model, the decoherence factor is expected to take the form as a Gaussian-exponential convolution. On the other hand, there are exceptions for these conditions as well. For instance, one can get a strict exponential decay of the decoherence factor by manually tuning the energy spectrum of the bath in to an exponential form \cite{Cucchietti2005-yh}. In the following discussion of the quantum Brownian motion, we will see another example, where at low temperatures, the non-Gaussian nature of the edge of the environment spectrum modifies the general behavior of the decoherence factor.

\begin{figure}[t!]
    \centering
    \includegraphics[width=0.55\textwidth]{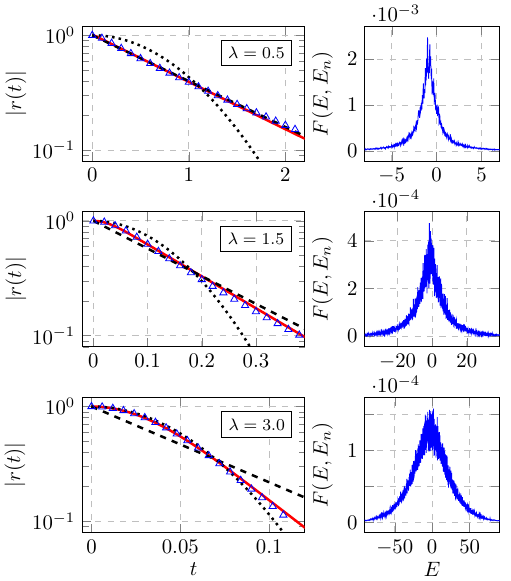}
    \caption{Decay of the decoherence factor at various coupling strengths $\lambda$ in semi-log scale, with the corresponding overlap function (second column). The initial state of the environment is prepared as a pure energy eigenstate in the middle of the spectrum. Triangles are the numerical data, with the corresponding convolution functions (\ref{eq:convolution}) plotted in red curves. To guide the eyes, we also plotted the best fit to Gaussian (dotted) and exponential (dashed) functions.}
    \label{fig:decay}
\end{figure}

To illustrate the above results, we numerically simulate a specific bath term consisting many interacting spin-$1/2$ particles. More precisely, the linear coupling term and the bath internal Hamiltonian in Eq.~(\ref{eq:Hamiltonian}) are;
\begin{eqnarray}\label{eq:randomqubit}
        H_\mathcal{I} = \sum_i a_i\sigma^z_{\mathcal{E},i},\quad
        H_\mathcal{E} = \sum_\alpha\sum_{i\ne j} b^\alpha_{ij} \sigma^\alpha_{\mathcal{E},i}\sigma^\alpha_{\mathcal{E},j},
\end{eqnarray}
where $\alpha=x,y,z$, and $\sigma^\alpha_{\mathcal{E},i}$ are Pauli operators vectors for the environment spins while $a_i$ and $b^\alpha_{ij}$ are random numbers drawn from standard normal distribution.
\begin{figure}[t!]
    \centering
    \includegraphics[width=0.55\textwidth]{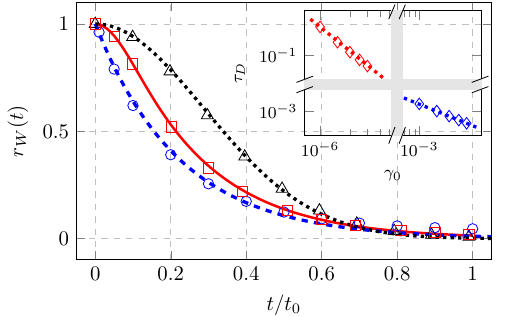}
    \caption{Decoherence in the quantum Brownian motion at various values of $\gamma_0$. The curves are plotted on the same time scale with the aid of a numerical re-scaling parameter $t_0$. Black triangles, blue circles, and red diamonds correspond to $\{\gamma_0, t_0\} = \{10^{-3},3.0\times 10^{-3}\}$, $\{10^{-6},1.0\}$ and $\{3\times 10^{-5},3.3\times 10^{-2}\}$, and are fitted to Gaussian, exponential, and convolution functions, respectively. The simulations are performed at temperature $T=2.5\times 10^4$ with a high-energy cutoff $\Lambda=500$. Inset: Decoherence time $\tau_D$ (in log-log scale) in the Gaussian (blue diamonds) and exponential (red diamonds) regimes, which scales as $\sim 1/\gamma_0^{1/2}$ (blue dotted) and $\sim 1/\gamma_0$ (red dotted), respectively.}
    \label{fig:QBmotion}
\end{figure}

In Fig.~\ref{fig:strength} (inset), the spectral density is shown to be well described by a Gaussian distribution. The claim for the Lorentzian distribution of the overlap is also justified. Note that the asymptotic tail of the overlap becomes exponential. This agrees with Wigner's derivation \cite{Wigner1955-jp,Wigner1957-cv} for the tail of the strength function of banded matrices. The matrix of the physical perturbation represented in the basis of the unperturbed Hamiltonian is a banded matrix, i.e., its off-diagonal elements typically decay exponentially fast with the distance from the diagonal \cite{Deutsch2018-ri}.

To visualize the change of the time dependence of the decoherence factor with the growth of the coupling strength (and hence, with the width of the overlap), here we focus on the case where the environment is initially prepared in an energy eigenstate. The results can be generalized to a thermal environment by averaging over a thermal distribution. This will be addressed in the study of quantum Brownian motion.
Figure~\ref{fig:decay} depicts the decoherence factor at various coupling strengths, demonstrating the transition between exponential and Gaussian decays, with a crossover well-described by their convolution.

\section{Quantum Brownian motion}

Encouraged by the above study of the spin-bath model, we expect that the convolution paradigm is robust and applies to other decoherence problems. We now test this conjecture with the familiar and extensively studied quantum Brownian motion (QBM) model \cite{Feynman1963,Caldeira1983,Unruh1989,Hu1992-jl}. Surprisingly, the possibility of the exponential to Gaussian transition has been so far overlooked. 

QBM describes a single harmonic oscillator with a unit mass and a unit bare frequency, linearly interacting with a thermal bath of non-interacting harmonic oscillators. The interaction Hamiltonian between the system oscillator and the bath oscillators is
\begin{equation}
    H_\mathcal{I} = -\sum_k c_k\hat{x}_\mathcal{S}\hat{x}^k_\mathcal{B},
\end{equation}
where $c_k$ are the coupling strength. This model can be solved exactly \cite{Hu1992-jl,Paz1993-il}. 

It is well-known that, to study the reduced dynamics of the system, one can cast the effect of the bath oscillators into a single function, the spectral density function
\begin{equation}\label{eq:spectralden1}
    I(\omega) = \sum_k \frac{c^2_k}{2m_k\omega_k}\delta(\omega-\omega_k),
\end{equation}
where $m_k$ and $\omega_k$ are the masses and frequencies of the bath oscillators. Here, we consider a Ohmic environment, as it is widely studied in the literature. Its spectral density function is given by
\begin{equation}
    I(\omega) = \frac{2 \gamma_0}{\pi} \omega \exp{\left(-\frac{\omega^2}{\Lambda^2}\right)}.
\end{equation}
Above, $\Lambda$ is a high-energy cutoff, and $\gamma_0$ quantifies the interaction strength. (Compared to Eq.~(\ref{eq:spectralden1}), it can be seen that $\gamma_0$ is quadratic in the linear coupling strength between the system and environment oscillators). We provide detailed solution of the QBM in the appendix, where it is clear how the spectral density function enters the dynamics of the system oscillator.

We assume the system and the bath are initially uncorrelated. The bath is in a thermal state with temperature $T$, and the system oscillator is prepared in a ``Schr\"odinger cat'' superposition of two Gaussian wavepackets located at different positions, i.e., $\psi(t=0) = \psi_1 + \psi_2 $, with
\begin{eqnarray}
    \psi_{1,2}(x) = \mathcal{N} \exp{\left(-\frac{(x\mp x_0)^2}{2\delta^2}\right)},
\end{eqnarray}
where $\mathcal{N}$ is a normalization factor. Due to couplings to the bath, the system oscillator evolves into a mixed state, which can be written as
\begin{eqnarray}\label{eq:reduce}
    \rho(t) = 
    \rho_1(t) + \rho_2(t) + \rho_{\rm int}(t),
\end{eqnarray}
where $\rho_{1,2}$ are the reduced states resulting from the evolution of $\psi_{1,2}$ alone, and $\rho_{\rm int}$ is the interference term between them \cite{Zurek1986,Zurek1991}. 

Decoherence of the system is reflected in the decay of $\rho_{\rm int}$. To quantify this, we transform each term of the reduced state (\ref{eq:reduce}) into a corresponding Wigner representation in the phase space, $W_{1,2}$ and $W_{\rm int}$, and study the decay of the peak-to-peak ratio between the Wigner functions (see Appendix C.1):
\begin{eqnarray}
                r_W(t) \equiv \frac{1}{2} \frac{W_{\rm int}|_{\rm peak}}{W_{1(\rm{or}\ 2)}|_{\rm peak}}.
\end{eqnarray}
This quantity is known to be a good measure of decoherence for QBM \cite{Paz1993-il}.
Figure~\ref{fig:QBmotion} depicts the decay of $r_W$ at various values of $\gamma_0$. Both the exponential and Gaussian regime are accurately described by the convolution paradigm.

For QBM, we define the decoherence time $\tau_D$ through $r_W(\tau_D) = 1/e$. Since $\gamma_0$ in the spectral density is quadratic in the coupling strength between the system and environment oscillators \cite{Hu1992-jl}, the decoherence time $\tau_D$ is thus expected to scale as $\sim 1/\gamma_0$ and $\sim 1/\gamma_0^{1/2}$, in the exponential and Gaussian regime, respectively. Both of these scaling behaviors are verified in the inset of Fig.~\ref{fig:QBmotion}. We present in the appendix the full solution of the model and the rationale for the convolution description.

\begin{figure}[t!]
    \centering
    \includegraphics[width=0.8\textwidth]{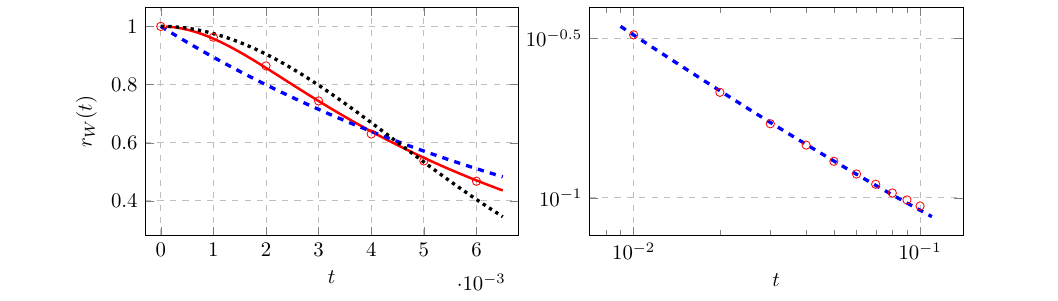}
    \caption{Decoherence in the quantum Brownian motion at zero temperature with $\gamma_0=0.01$, and frequency cut-off $\Lambda=500$. Red circles are the numerical data of $r_W$. Left: At early times, decay of $r_W$ is well-described by the convolution function (red solid curve). To guide the eyes, we also plotted the best Gaussian (black dotted) and exponential (blue dashed) fit. Right (log-log scale): At late times, decay of $r_W$ is suppressed by a power law. The blue dashed line is fitted to $0.017t^{-0.63}+0.02$.}
    \label{fig:QBmotionT0}
\end{figure}

As discussed before, the convolution behavior of the decoherence factor is accurate when the spectrum of the environment is well described by a Gaussian distribution, and the overlap is given by the Lorentzian-like distribution (\ref{eq:spread}). These conditions may not be fulfilled when the state of the environment is close to the edge of the spectrum, e.g., the temperature is extremely low. In this case, the overlap cannot extend below the ground state energy. This low-energy cutoff may modify the shape of the overlap, and hence reduce the accuracy of our analysis in terms of convolution functions. To test this situation, we simulated $r_W(t)$ at zero temperature, shown in Fig.~\ref{fig:QBmotionT0}. It can be seen that at early times, $r_W(t)$ can still be accurately described by the convolution function (Fig.~\ref{fig:QBmotionT0} left). This further demonstrated the robustness of the convolution paradigm. The low energy cutoff of the overlap mainly affects the long time behavior of its Fourier transform. Indeed, at long times, as shown in Fig.~\ref{fig:QBmotionT0} (right), the decay of $r_W(t)$ is suppressed by a power-law.

It is worth stressing that, the range of $\gamma_0$ shown in Fig.~\ref{fig:QBmotion} (main plot) is small, so that the time dependence appears to be Markovian \cite{Paz1993-il}. We have developed an exact numerical treatment, aimed at exploring of the exponential-to-Gaussian transition through a wide range of coupling strength. We have focused on the Markovian regime in this paper because the exponential-to-Gaussian crossover turned out to occur already in this regime. The question is nevertheless whether the results presented above still hold when the coupling is strong. 
Our theory predicts that, for coupling strengths stronger than that in the crossover regime, we should observe Gaussian decays of the decoherence factor. The exact numerical treatment allows us to confirm that. As expected, we observed nearly perfect Gaussian decay for larger couplings (blue line in the inset of Fig.~\ref{fig:QBmotion}). 

Now the puzzling fact is that our main result predicts exponential-to-Gaussian transition from weak to strong couplings, respectively, with a crossover in between. However, for the QBM the crossovers have all been observed already in the weak coupling (Markovian) regime. This looks like a disagreement. The answer to this apparent problem is that the QBM dynamics do not quite fit into the spin-bath model. The appearance of the convolution and of the exponential-to-Gaussian transition is due to the presence of an effective ``echo’’ operator (\ref{eq:decoherencefactor0}) (detailed in Appendix C.3). As a result, this transition is not completely determined by the coupling strength $\gamma_0$. Rather, the effective coupling strength is set by $\gamma_0$ together with the relevant energy scale of the system. Let us briefly discuss the mechanism here. Suppose the coupling term between the system $\mathcal{S}$ and the bath $\mathcal{B}$ is
    $H_{\mathcal{I}} = H_{\mathcal{I}}^{\mathcal{S}} \otimes H_{\mathcal{I}}^{\mathcal{B}}$.
In the truncated Hilbert space below the high energy cut-off, we can write the eigenvalues of $H_\mathcal{I}^S$ as $\{\lambda_k\}_{k=1,2...}$. As shown in the appendix, $\lambda_k$ serves as the effective ``coupling strength'' that controls the exponential-to-Gaussian transition. Now, $\lambda_k$ becomes proportional to the true coupling strength of the QBM, $\gamma_0^2$. It also depends on the energy scale of the system, so that at high temperatures, $\lambda_k$ can extend to large values. Therefore, even though $\gamma_0$ is in the Markovian regime, the effective strength $\lambda_k$ can be in the strong regime. 

The above argument also suggests that, at lower temperatures, we should expect that the crossover regime happens at larger values of $\gamma_0$. This is indeed confirmed by comparing Figs.~\ref{fig:QBmotion} and \ref{fig:QBmotionT0}. Quantitative analysis of this interplay deserves further investigation.

\section{Discussion}

We have provided a unified and robust description of the interplay between the exponential to Gaussian decoherence in terms of the convolution functions. The relative import of these two types of decoherence is controlled by the interaction strength between the system and the environment as was seen in the two paradigmatic decoherence models, the spin-bath and the quantum Brownian motion. 

The convolution description of the decoherence factor is explicitly derived using the standard spin-bath model, and therefore is not expected to work in any circumstance of decoherence. However, because of its simplicity, it is robust and can be used to model a broad class of systems. Our result straightforwardly applies to such scenarios. The quantum Brownian motion (QBM) does not fit into the paradigm of the spin-bath model. However, as elaborated in the previous section, the QBM exhibits similar structure that is responsible for the convolution description of the decoherence factor, namely, the appearance of the echo operator in (\ref{eq:decoherencefactor0}). This is responsible for the observed accurate description of decoherence using convolution functions in the QBM model. Moreover, we expect that the convolution paradigm applies to a broader context than decoherence, whenever the same imperfect echo process is involved. This may explain the Gaussian to exponential transition in information scrambling \cite{Yan2020-ff}, decay processes \cite{Usaj1998-ka,Flambaum2001-xi}, as well as recent experiments \cite{Sanchez2020-wx} showing similar crossover behaviors (although study of the convolution paradigm in fields such as nuclear physics where LDOS has very different origins than in decoherence is beyond the scope of our discussion).

Decoherence is the main obstacle in implementing quantum computing. The time dependence of the decoherence factor will affect the error correction strategy (e.g., selection of the time intervals between the error correcting operations). We hope our result will provide valuable input into the hardware design in overcoming decoherence. 

\section*{Acknowledgements}
The authors acknowledge Nikolai Sinitsyn and Lukasz Cincio for helpful discussion, Jess Riedel and Michael Zwolak for constructive comments on the manuscript. This research was supported by the U.S Department of Energy under the LDRD program in Los Alamos. B.Y. also acknowledges support from U.S. Department of Energy, Office of Science, Basic Energy Sciences, Materials Sciences and Engineering Division, Condensed Matter Theory Program, and the Center for Nonlinear Studies. W.H.Z. acknowledges partial support by the Foundational Questions Institute’s Consciousness in the Physical World program, in partnership with the Fetzer Franklin Fund.

\appendix

\section{Effective perturbation}

The echo operator in the decoherence factor, Eq.~(\ref{eq:decoherencefactor0}) in the main text, is written in term of an effective perturbation $H_\mathcal{P}$, i.e.,
\begin{eqnarray}       
e^{i(H_\mathcal{E}+ H_\mathcal{I})t}e^{-i( H_\mathcal{E}- H_\mathcal{I})t} = e^{i( H_\mathcal{E}+H_\mathcal{P})t}e^{-i H_\mathcal{E}t}.
\end{eqnarray}
In practice, the interaction $H_\mathcal{I}$, as well as the effective perturbation $H_\mathcal{P}$, are much smaller compared to the environment Hamiltonian $H_\mathcal{E}$. This allows to expand echo operator as a power series. Intuitively, the effective perturbation $H_\mathcal{P}$ should be twice as large as the interaction $H_\mathcal{I}$, as it only appears in the forward evolution loop. This can be verified by expanding the logarithm of the echo operator. Using the Baker-Campbell-Hausdorff formula, up to first order, we have
\begin{eqnarray}
       &\log{\left(e^{i(H_\mathcal{E}+ H_\mathcal{I})t}e^{-i( H_\mathcal{E}- H_\mathcal{I})t}\right)}\\
        =& 2i t H_\mathcal{I} + \frac{1}{2}t^2 \left[ 2H_\mathcal{I}, H_\mathcal{E} \right] + \frac{1}{12}2it^3 \left[H_\mathcal{E}, \left[ 2H_\mathcal{I}, H_\mathcal{E} \right]\right] \\
        & \quad\quad\quad -\frac{1}{24}2t^4 \left[H_\mathcal{E}\left[H_\mathcal{E},2 \left[ 2H_\mathcal{I}, H_\mathcal{E} \right]\right]\right] + \cdots\\
\end{eqnarray}
and
\begin{eqnarray}
        &\log{\left(e^{i( H_\mathcal{E}+H_\mathcal{P})t}e^{-i H_\mathcal{E}t}\right)}\\
        =& itH_\mathcal{P} + \frac{1}{2}t^2 \left[ H_\mathcal{P}, H_\mathcal{E} \right] + \frac{1}{12}2it^3 \left[H_\mathcal{E}, \left[ H_\mathcal{I}, H_\mathcal{P} \right]\right] \\
        &\quad\quad\quad -\frac{1}{24}2t^4 \left[H_\mathcal{E}\left[H_\mathcal{E}, \left[ H_\mathcal{P}, H_\mathcal{E} \right]\right]\right] + \cdots
\end{eqnarray}
Hence, to the first order, $H_\mathcal{P} = 2H_\mathcal{I}$.

\section{The convolution function}

For the standard Fourier transform $\hat{\mathcal{F}}$,
\begin{eqnarray}
    \hat{\mathcal{F}}[f(x)]\equiv \int dx\ f(x)e^{-2\pi i xt},
\end{eqnarray}
the convolution theorem asserts
\begin{eqnarray}
    \hat{\mathcal{F}}[fg] = \hat{\mathcal{F}}[f]*\hat{\mathcal{F}}[g],
\end{eqnarray}
where the convolution $*$ is defined as
\begin{eqnarray}
    f*g \equiv \int_{-\infty}^\infty d\tau\ f(\tau)g(x-\tau).
\end{eqnarray}

For convenience, we adapt the Fourier transform for the angular frequency $2\pi x$, namely,
\begin{eqnarray}
    \hat{\mathcal{F}}[f(x)]\equiv \int dx\ f(x)e^{- i xt},
\end{eqnarray}
with the corresponding inverse Fourier transform normalized by a pre-factor $1/2\pi$. Under this convention, the convolution theorem reads

\begin{eqnarray}
        \hat{\mathcal{F}}[fg] &= 2\pi \int dx\ e^{- 2\pi i xt} f(2\pi x)g(2\pi x),\\
        &= 2\pi \int dx\ e^{- i 2\pi xt} f(2\pi x) * \int dx\ e^{- i 2\pi xt} g(2\pi x)\\
        &= \frac{1}{2\pi}\hat{\mathcal{F}}[f(x)]\hat{\mathcal{F}}[g(x)].
\end{eqnarray}
This allows to evaluate the decoherence factor in the main text as

\begin{eqnarray}
        r(t) &= e^{iE_nt} \int d E \ e^{-iEt} F(E,E_n) \eta(E)\\
    &  = \frac{1}{2\pi} e^{iE_nt}\hat{\mathcal{F}}\left[F(E,E_n)\right]*\hat{\mathcal{F}}\left[\eta(E)\right].
\end{eqnarray}

Assuming that the global spectrum density has a Gaussian shape with a band-width (standard deviation) $\sigma$, its Fourier transform is a Gaussian decay 
\begin{eqnarray}
    \hat{\mathcal{F}}\left[\eta(E)\right] = e^{-\sigma^2t^2/2}.
\end{eqnarray}
As derived in the main text, the overlap takes the form
\begin{eqnarray}
        F(E,E_n) = \frac{V^2}{(E-E_n+E_r)^2+(\Gamma_{\rm eff}/2)^2},
\end{eqnarray}
with an effective width $\Gamma_{\rm eff}$ and a small shift $E_r$ of the peak position $E_n$. Its Fourier transform is given by
\begin{eqnarray}
    \hat{\mathcal{F}}\left[F(E,E_n)\right] \propto e^{-i(E_n-E_r)t} e^{-\Gamma_{\rm eff}|t|/2}.
\end{eqnarray}
Here we omitted the normalization factor since the decoherence factor always start from unity. 

Evaluating the convolution of the above two Fourier transforms of the overlap and the spectral density, we arrive at the final expression of the decoherence factor
\begin{eqnarray}
    r(t) & \propto \  e^{-\Gamma |t|/2}*e^{-\sigma^2 t^2/2}\\
    & = \int\  d\tau\ e^{-\Gamma |\tau|/2}e^{-\sigma^2 (t-\tau)^2/2}\\
        & \propto \  e^{-\Gamma t/2} \rm{Erfc}\left(\frac{\Gamma/2-\sigma^2 t}{\sqrt{2}\sigma}\right)+ e^{\Gamma t/2} \rm{Erfc} \left(\frac{\Gamma/2+\sigma^2 t}{\sqrt{2}\sigma}\right),
\end{eqnarray}
Note that the above solution gives the unnormalized magnitude of the decoherence factor, which starts from unity at the initial time. The decoherence factor also has a residue oscillation $e^{-iE_rt}$ induced by the small shift $E_r$ of the resonance peak of the overlap function.

\section{Quantum Brownian motion}

Quantum Brownian (QBM) has been extensively studied in the literature. In this section, we present the model, and, to make the paper self-contained, collect the essential ingredients for solving it. We then discuss the convolution function description for decoherence of QBM.

\subsection{The model}

The model contains a one-dimensional harmonic oscillator (the system) with unit mass and bare frequency $\Omega_0$, and a collection of harmonic oscillators (the bath) with mass $\{m_k\}$ and frequency $\{\omega_k\}$. The system oscillator interacts linearly with the bath oscillators through
\begin{eqnarray}
    H_\mathcal{I} =  - \sum_k c_k \hat{x}_\mathcal{S}\hat{x}_\mathcal{B}^k,
\end{eqnarray}
with coupling strengths $\{c_k\}$. It is well-known that the effect of the bath oscillators on the reduced dynamics of the system oscillator can be fully characterized with a spectral density function, defined as
\begin{eqnarray}
    I(\omega) = \sum_k \frac{c^2_k}{2m_k\omega_k}\delta(\omega-\omega_k).
\end{eqnarray}
Here, we consider the particular form of the spectral density describing the Ohmic environment, as is widely used in the literature:
\begin{eqnarray}
    I(\omega) = \frac{2 \gamma_0}{\pi} \omega \exp{\left(-\frac{\omega^2}{\Lambda^2}\right)},
\end{eqnarray}
where $\Lambda$ is a high energy cut-off, and $\gamma_0$ is a constant that quantifies the linear coupling strength (it is quadratic in the coupling strength).

As considered in Ref.~\cite{Paz1993-il}, we assume the bath is prepared in a thermal state with temperature $T$, and the initial state of the system oscillator is in a superposition of two Gaussian wave packets,
\begin{eqnarray}
    \psi(x,t=0)=\psi_1(x) + \psi_2(x),
\end{eqnarray}
where $\psi_{1,2}$ are Gaussian wavefunctions localized at separated locations, i.e.,
\begin{eqnarray}
    \psi_{1,2} = \mathcal{N} \exp{\left( -\frac{(x \mp x_0)^2}{2\delta^2} \right)},
\end{eqnarray}
with $\mathcal{N}$ a normalization constant.
Due to coupling to the bath of oscillators, the state of the system oscillator will become mixed during time evolution, which can be factorized into
\begin{eqnarray}\label{appeq:reduce}
    \rho(t) = \rho_1(t) + \rho_2(t) + \rho_{\rm int}(t).
\end{eqnarray}
Here $\rho_{1,2}$ are the reduced density matrices from the evolution of $\psi_{1,2}$, and $\rho_{\rm int}$ is the interference part. The state can be visualized using the Wigner function in the phase space, defined for the density matrix as
\begin{eqnarray}
    W(x,p)\equiv \int \frac{dz}{2\pi}e^{ipz}\rho(x-z/2,x+z/2).
\end{eqnarray}
Represented as a Winger function, the reduced density matrix (\ref{appeq:reduce}) has three corresponding terms
\begin{eqnarray}\label{appeq:reducewigner}
    W(t) = W_1(t) + W_2(t) + W_{\rm int}(t).
\end{eqnarray}
As shown in Ref.~\cite{Paz1993-il}, $W_{1,2}$ remain as separated Gaussians in the phase space, while the interference term $W_{\rm int}$ decays in time. To quantify the decay of interference, we study the ratio between the peak values of $W_{\rm int}$ and $W_{1(2)}$:
\begin{eqnarray}\label{appeq:decofactor}
                r_W(t) \equiv \frac{1}{2} \frac{W_{\rm int}|_{\rm peak}}{W_{1(\rm{or}\ 2)}|_{\rm peak}},
\end{eqnarray}
This quantity has been demonstrated \cite{Paz1993-il} to be a good measure for decoherence. 

\subsection{The solution}
In this section, we collect the essential results of the solution that allow us to perform numerical simulations. More details of the derivations can be found in Refs.~\cite{Hu1992-jl,Paz1993-il}.

The total reduced density matrix (\ref{appeq:reduce}) can be solved using path integral \cite{Hu1992-jl}
\begin{eqnarray}
    \rho(x,y,t) = \int\int dx'dy' J(x,y,t;x',y',t')\rho(x',y',t'),
\end{eqnarray}
where the propagator, written in terms of the variables $X=x+y$ and $Y=x-y$, is given by
\begin{eqnarray}
    & J(X,Y,t;X',Y',t')\\
    = & \frac{b_3}{2\pi}\exp\left({-a_{11}Y^2 -a_{12}YY' - a_{22}Y'^2}\right)\\
    &\quad\quad\times \exp\left({ib_1XY + ib_2X'Y -ib_3XY' -ib_4X'Y' }\right).
\end{eqnarray}
Here, the functions $a_{ij}(t)$ and $b_{i}(t)$ can be determined by the equations
\begin{eqnarray}
    &\ddot{u}(s) + \Omega_0^2 u(s) + 2\int_0^s ds'\ \eta(s-s')u(s') = 0,\\
    &2b_1(t) = \dot{u}_2(t),\quad    2b_3(t) = \dot{u}_2(0),\\
    &2b_2(t) = \dot{u}_1(t),\quad   2b_4(t) = \dot{u}_1(0),\\
    &a_{ij} = \frac{1}{1+\delta_{ij}} \int_0^\infty\int_0^\infty dsds'\ u_i(s)u_j(s')\mu(s-s'),
\end{eqnarray}
with boundary conditions $u_1(0)=u_2(t)=1$ and $u_1(t)=u_2(0)=0$, and $\mu(s), \eta(s)$ are the noise and dissipation kernels determined by the spectral density:
\begin{eqnarray}
              \mu(s) &= \int_0^\infty d\omega \ I(\omega)\coth\left(\frac{\omega}{2T}\right)\cos\left(\omega s\right), \\
              \eta(s) &= -\int_0^\infty d\omega\ I(\omega)\sin\left(\omega s\right). 
\end{eqnarray}
We solve the above integrodifferential equation for $u(t)$ with a shooting method incorporating the imposed boundary conditions. With these solutions, we can then compute the peak-to-peak ratio (\ref{appeq:decofactor}) between the Wigner functions:
\begin{eqnarray}
                r_W(t) & =  \frac{1}{2} \frac{W_{\rm int}|_{\rm peak}}{W_{1(\rm{or}\ 2)}|_{\rm peak}}\\
    &=\exp{\left( -x_0^2/\delta^2 + \kappa_p^2/\delta_2^2 +\delta_1^2\kappa_x^2 \right)},
\end{eqnarray}
where 
\begin{eqnarray}
\delta_1^2 =& \left( a_{22} + \frac{1}{4\delta^2 + \delta^2b_4^2} \right)b_3^{-2},\\
\delta_2^2 =& \frac{1}{4}\left[ a_{11} +\delta^2b_2^2 - \frac{1}{4\delta_1^2} \left( \frac{a_{12-2\delta^2b_2b_4}}{b_3}^2 \right) \right]^{-1}, \\
\kappa_x =& \frac{x_0}{2\delta_1^2\delta^2b_3},\quad \kappa_p =  \frac{x_0\delta_2^2(a_{12}-2\delta^2b_2b_4)}{2\delta^2\delta_1^2b_3^2} .
\end{eqnarray}
Here, the peak values for $W_{1,2}$ are the same, so we do not distinguish between them.

\subsection{Convolution in QBM}

As shown in the main text, the convolution function fits the decoherence factor of QBM very well. In this section, we explore the mechanism behind this. Instead of solving the reduced dynamics of the system harmonic oscillator alone, here we need to include the bath degrees of freedom into consideration. 

The bath is prepared in a thermal state. However, for simplicity, in the following discussion we assume the bath is initially in a pure energy eigenstate. The result can be generalized to thermal states by doing a simple thermal average. Under this condition, the initial total wavefunction of the system oscillator and the bath is
\begin{eqnarray}
    \Psi(t= 0) =(\psi_1+\psi_2)\otimes\psi_\mathcal{B},
\end{eqnarray}
which results in two branches of the total wavefunction at any time instant $t$, i.e.,
\begin{eqnarray}
    \Psi(t) = \Psi_{1,\mathcal{B}} + \Psi_{2,\mathcal{B}}.
\end{eqnarray}
The reduced states of the system oscillator from each branch, e.g., $\rho_{1,2} = \rm{Tr} |\Psi_{1,2,\mathcal{B}}\rangle\langle \Psi_{1(2),\mathcal{B}}|$,  are mixed states, whose Wigner representations (the first two terms in Eq.~(\ref{appeq:reducewigner})) in the phase space take Gaussian forms. As shown in Ref.~\cite{Paz1993-il}, during time evolution, these two Gaussians are always separated and are approximately orthogonal to each other, as are the system's reduced density matrices $\rho_{1,2}(t)$ -- they are (approximately) supported on orthogonal subspaces of the total Hilbert space. Therefore, we can decompose the total wavefunction at any time instant into:
\begin{eqnarray}\label{appeq:expansion}
    \Psi(t) \approx \sum_k^N \psi_1^k(t)\otimes\psi^k_{\mathcal{B}1}(t) + \sum_k^N \psi_2^k(t)\otimes\psi^k_{\mathcal{B}2}(t).
\end{eqnarray}
In this expansion, the basis states for the first branch, $\{\psi_1^k(t)\}_k$, are orthogonal to those of the second branch, $\{\psi_2^k(t)\}_k$. Here, we truncate the infinite dimensional Hilbert space to a finite (but large) $N$-dimensional Hilbert space within the relevant energy scale. By choosing a sufficiently large $N$, this can approximate the exact wave function to any degree of accuracy.

The above assumptions are specifically made for the QBM model and are justified by the known solutions.  From now on, we will keep the discussion in a more formal and abstract manner beyond the QBM model. We write the linear interaction in a generic form
\begin{eqnarray}
    H_\mathcal{I} = H_\mathcal{I}^\mathcal{S} \otimes H_\mathcal{I}^\mathcal{B}.
\end{eqnarray}
As a matter of convention, we also choose the basis $\{\psi_{1,2}^k(t)\}_k$ as an eigenbasis of the interaction Hamiltonian $H_\mathcal{I}^\mathcal{S}$ -- this is always possible since for any basis of the truncated Hilbert space we can diagonalize $H_\mathcal{I}^\mathcal{S}$ and use the resulting eigenbasis to expand the wavefunction into the form (\ref{appeq:expansion}). The corresponding eigenvalues of the basis are denoted as $\{\lambda_{1(2)}^k(t)\}_k$.

As introduced in the previous sections, the reduced state of the system oscillator can be factorized into
\begin{eqnarray}
    \rho(t) = \rho_1(t) + \rho_2(t) + \rho_{\rm int}(t),
\end{eqnarray}
where $\rho_{1}$ ($\rho_{2}$) is reduced from the first (second) branch of the total wavefunction, and $\rho_{\rm int}$ comes from the interference part. In our $2N$-dimensional truncated total Hilbert space, $\rho_{\rm int}$ has matrix element
\begin{eqnarray}\label{appeq:rmatrix}
    \rho_{\rm int}^{1i,2j} = \rm{Tr} |\psi^{i}_{\mathcal{B}1}(t)\rangle\langle\psi^{j}_{\mathcal{B}2}(t)|.
\end{eqnarray}
With this, we can build up the evolution trajectories for the bath wavefunctions $\{\psi_{\mathcal{B}1}^k(t)\}_k$ and $\{\psi_{\mathcal{B}2}^k(t)\}_k$. Consider the evolution of the total wavefunction from the given state $\Psi(t)$ at time $t$ to time $t+\Delta t$ with an infinitesimal $\Delta t$. The evolution reads 
\begin{eqnarray}
        \Psi(t+ \Delta t) & = e^{-iH\Delta t} \Psi(t)\\
        & = \sum_k^N \psi_1^k(t)\otimes e^{-i\left(H_0+\lambda_1^k(t) H_I^\mathcal{B}\right)\Delta t}\psi^k_{\mathcal{B}1}(t) \\
        & \quad\quad + \sum_k^N \psi_2^k(t)\otimes e^{-i\left(H_0+\lambda_2^k(t) H_I^\mathcal{B}\right)\Delta t} \psi^k_{\mathcal{B}2}(t),
\end{eqnarray}
where $H_0$ is the internal Hamiltonian (non-interacting part) of the total system. Hence, we conclude that the trajectory $\psi_{\mathcal{B}1}^k(t)$ (and similarly for the second branch $\psi_{\mathcal{B}2}^k(t)$) can be interpreted as an evolution with a time-dependent Hamiltonian, i.e.,
\begin{eqnarray}
    \psi_{\mathcal{B}1}^k(t) = \overleftarrow{\mathcal{P}} e^{-i\int d\tau \left(H_0+\lambda_1^k(\tau) H_I^\mathcal{B}\right)\tau}\psi^k_{\mathcal{B}1}(0),
\end{eqnarray}
where the time ordering $\overleftarrow{\mathcal{P}}$ is imposed. Plugging the above equation into Eq.~(\ref{appeq:rmatrix}), we get the matrix element of the interference part of the reduced density matrix,
\begin{eqnarray}
            \rho_{\rm int}^{1i,2j} & = \rm{Tr}\left[ e^{-i\int d\tau \left(H_0+\lambda_1^i(\tau) H_I\right)\tau}|\psi^{i}_{\mathcal{B}1}(0)\rangle \langle \psi^{j}_{\mathcal{B}2}(0)|e^{i\int d\tau \left(H_0+\lambda_2^j(\tau) H_I\right)\tau} \right]\\
            & =\langle \psi^{j}_{\mathcal{B}2}(0)|e^{i\int d\tau \left(H_0+\lambda_2^j(\tau) H_I\right)\tau} e^{-i\int d\tau \left(H_0+\lambda_1^i(\tau) H_I\right)\tau}|\psi^{i}_{\mathcal{B}1}(0)\rangle\\
            & =\langle \psi_{\mathcal{B}}|e^{i\int d\tau \left(H_0+\lambda_2^j(\tau) H_I\right)\tau} e^{-i\int d\tau \left(H_0+\lambda_1^i(\tau) H_I\right)\tau}|\psi_{\mathcal{B}}\rangle.
\end{eqnarray}
Note that in the above equation we used a boundary condition imposed by the initial state of the bath, i.e., $\psi_{\mathcal{B}1,2}^k(0) = \psi_\mathcal{B}$. We also omitted the time ordering operator, but the integral is time ordered and is interpreted as an evolution generated by the Hamiltonian $H_0$ with small time-dependent perturbations $\lambda_{1,2}(t)H_\mathcal{I}^\mathcal{B}$. 

Hence, the interference part of the reduced density matrix is related to the echo operators, which then contribute to the convolution function in the same manner as in the qubit model discussed in the main text. It is worth emphasizing that the above derivation is by no means a method for solving the reduced density matrix from the original model Hamiltonian, since the effective perturbations $\lambda_{1,2}(t)$ have to be extracted from the solution (\ref{appeq:expansion}) known in the first place. The purpose for this transformation is to extract the structure in terms of the echo operators.

\bibliography{reference}

\end{document}